\documentclass[aps,prl,reprint,groupedaddress,showpacs]{revtex4-1}

\usepackage{hyperref}
\usepackage{amssymb,amsmath,amsfonts,mathtools}
\usepackage{epsfig}
\usepackage{enumerate}
\usepackage{subfigure}
\usepackage[bulletsep]{collref}
\usepackage{color}

\newcommand{\half}{\frac{1}{2}}

\def\s{\sigma}

\def\l{\lambda}

\newcommand{\CL}{\mathcal{L}}

\newcommand{\DIV}[1][\normalsize]{\,\mbox{div}\,}

\def\[{\begin{equation}}
\def\]{\end{equation}}
\newcommand{\be}{\begin{eqnarray}}
\newcommand{\ee}{\end{eqnarray}}
\newcommand{\nn}{\nonumber}

\begin{document}


\title{Magnetic oscillations in a holographic liquid}


\author{V. Giangreco M. Puletti$^{a}$,~S.~Nowling$^{b}$,~ L.~Thorlacius$^{a,c}$,~ T.~Zingg$^{d}$}
\affiliation{$^{a}$ University of Iceland, Science Institute, Dunhaga 3, 107 Reykjavik, Iceland
\\
$^{b}$ Jamestown Community College, 525 Falconer St, Jamestown, NY 14701, USA
\\
$^{c}$ The Oskar Klein Centre for Cosmoparticle Physics, Department of Physics, Stockholm University, 
AlbaNova University Centre, 106 91 Stockholm, Sweden\\
$^{d}$
Helsinki Institute of Physics\\P.O. Box 64, FI-00014 University of Helsinki, Finland
 }



\begin{abstract}
We present a holographic perspective on magnetic oscillations in strongly correlated 
electron systems via a fluid of charged spin 1/2 particles outside a black brane in an 
asymptotically anti-de-Sitter spacetime. The resulting back-reaction on the spacetime 
geometry and bulk gauge field gives rise to magnetic oscillations in the dual field theory, 
which can be directly studied without introducing probe fermions, and which differ from 
those predicted by Fermi liquid theory.
\end{abstract}


\maketitle


\paragraph{Introduction:}
The de Haas -- van Alphen effect~\cite{1930dHvA} refers to quantum oscillations in the magnetization
as a function of $1/B$, present in metals at low temperature $T$ and 
strong magnetic field $B$. 
This phenomenon is generally associated with the Fermi surface and the observed 
oscillations are usually interpreted in terms of Fermi liquid theory and 
quasi-particles~\cite{Lifshitz-Kosevich}. The measurement of quantum oscillations is a 
standard tool for investigating the electronic structure of metallic systems and this extends to strange 
metallic phases, where fermion excitations are strongly correlated and the quasi-particle description 
breaks down. The observation of quantum oscillations, in combination 
with ARPES experiments, leads to the conclusion that low temperature physics in such systems 
is still governed by a Fermi surface, although the underlying physics is not fully understood, see
{\it e.g.}~\cite{0034-4885-75-10-102501,Sebastian28042011}, and may not conform to the standard 
Fermi liquid picture. In this paper, we provide a novel view of quantum oscillations that is not tied to 
a quasi-particle description but is instead based on a holographic representation of the strongly 
correlated electron system in terms of a dual gravitational model. 

In recent years, gauge/gravity duality~\cite{Maldacena:1997re,Gubser:1998bc,Witten:1998qj}, 
has been applied to model strongly coupled dynamics in various condensed matter systems including 
strange metals (for reviews see {\it e.g.}~\cite{Hartnoll:2011fn,Sachdev:2011wg}). 
Holographic systems at finite charge density exhibit interesting non-Fermi liquid behavior, revealed 
for instance in spectral functions of probe fermions~\cite{Liu:2009dm,Cubrovic:2009ye}. Here we 
develop a simple holographic model for strongly correlated fermions in a magnetic field and use 
it to study magnetic oscillations in an unconventional setting. The model involves a fluid of 
charged spin 1/2 particles outside a dyonic black brane in 3+1 dimensional anti-de Sitter (AdS)
spacetime. It takes into account the back-reaction due to charged bulk matter on the 
spacetime geometry and the bulk gauge field and thus extends the so-called electron star 
model~\cite{Hartnoll:2009ns,Hartnoll:2010gu} to include a magnetic field and finite temperature. 
We obtain oscillations in the magnetization directly from the bulk 
gravitational physics by incorporating Landau quantization into the charged fluid description. 
The method differs conceptually from earlier probe fermion computations of magnetic oscillations
in an electron star background~\cite{Gubankova:2010rc,Hartnoll:2011dm} and it predicts a 
dependence on the magnetic field and temperature of the oscillation amplitude that departs 
from Fermi liquid theory. 

\paragraph{The Model:}
Our approach is based on an extension of the electron star geometry,
developed in~\cite{Hartnoll:2009ns,Hartnoll:2010gu,Puletti:2010de,Hartnoll:2010ik}. For related 
work see also~\cite{deBoer:2009wk,Arsiwalla:2010bt}.
This geometry is obtained by coupling Einstein -- Maxwell theory to a charged perfect fluid,
of non-interacting fermions of mass $m$ and charge normalized to one in units of the Maxwell 
coupling constant $e$,
\be
\label{def_action_app}
S &=&	\frac{1}{2\kappa^2} \int d^4 x \sqrt{-g} \left( R- 2 \Lambda\right)	\nn	\\
  & & 		\;-\frac{1}{4 e^2}\int d^4 x \sqrt{-g} F_{\mu \nu} F^{\mu\nu}
			- \int d^4 x \sqrt{-g} \, \mathcal{L}_{\rm fl} \,, \qquad
\ee
where $\kappa^2=8\pi G_N$ is the gravitational coupling, the AdS length scale $L$ 
is given in terms of the negative cosmological constant $\Lambda=-3/L^2$, 
and $\kappa/L \ll 1$ corresponds to the classical gravity (large $N$) regime.
The bulk fermions are treated in a Thomas -- Fermi approximation, valid for model parameters satisfying
\be\label{esregime}
&& m L \gg 1\,, \qquad e^2\sim \frac{\kappa}{L} \ll 1\,.
\ee
In~\cite{Allais:2013lha, Medvedyeva:2013rpa} more refined computations involving holographic 
Fermi systems confirmed that the electron star qualitatively reproduces essential features of these models, 
even beyond its {\it a priori} regime of validity defined by equation \eqref{esregime}. 

Previous work on magnetic effects in holographic metals 
\cite{Denef:2009yy,Hartnoll:2009kk,Hartnoll:2010xj,Gubankova:2010rc,Blake:2012tp,Albash:2012ht,Gubankova:2013lca}
has not taken into account the full back-reaction on the geometry due to the presence of charged matter 
in a non-vanishing magnetic field at finite temperature. The semi-classical approximation used in the 
electron star construction, suitably generalized to finite $B$ and $T$,  also allows the back-reaction to be 
included in the $B \gg T$ regime, which is of primary interest to the study of magnetic oscillations. Including
the back-reaction provides direct access to the underlying strong coupling dynamics without having to 
introduce probe fermions.

We work in a 3+1 dimensional spacetime with local coordinates $(t,x,y,r)$, which asymptotes to 
$\mbox{AdS}_4$ and which is static, stationary and has translational symmetry orthogonal to the 
radial direction. At any given point in the spacetime the fluid is at rest in a local Lorentz frame 
$e^A_\mu$, and the fluid velocity is given by $u^\mu = e_0^\mu$.
The state of the charged fluid is completely determined by a local chemical potential and magnetic field,
\be
\label{localvariables}
\mu_{\rm loc}=A_\mu u^\mu\; , \qquad
H_{\rm loc}=  \, e^{[\mu}_1e^{\nu]}_2 F_{\mu\nu}	\; ,  \quad
\ee
where $A_\mu$ is a $U(1)$ gauge potential, and $F_{\mu\nu}$ the corresponding field 
strength tensor. This means that the charge density $\sigma$, the pressure $p$, 
and the magnetization density $\eta$ of the fluid depend only on $\mu_{\rm loc}$ and $H_{\rm loc}$.
An equation of state for the fluid is obtained as in~\cite{Hartnoll:2010gu}, except now the 
constituent dispersion relation is that of Dirac fermions in a magnetic field,
\be
\label{disp_rel}
E_\ell^2=	m^2 + k^2 + \big(2\ell +1\big) \gamma H_{\rm loc} \pm \gamma  H_{\rm loc}\,,
\ee
where $\gamma$ is a constant proportional to the gyromagnetic ratio of the fermions.
The index $\ell \geq 0$ labels Landau levels and the the last term on the right hand side 
is due to Zeeman splitting.
Assuming $1/N$ effects due to bulk thermalization can be neglected~\cite{Puletti:2010de,Hartnoll:2010ik}, 
the local pressure, charge and magnetization of the fluid are then obtained
by filling states up to the Fermi level given by $\mu_{\rm loc}$.
The number of occupied Landau levels is thus a local quantity,
\[
\label{levelcondition}
\ell_{\rm filled}=\left\lfloor \frac{ \mu_{\rm loc}^2 - m^2}{2 \gamma H_{\rm loc}}\right\rfloor \,,
\]
which is determined by the equations of motion and  
varies with radial position in the bulk geometries of interest. In particular, there is 
no fluid in regions where $\ell_{\rm filled}<0$.

The field variables only depend on the radial coordinate $r$ and
we choose the following parameterization for the metric,
\[\label{eq:metric}
ds^2 =\frac{L^2}{r^2} \left( 
-\frac{\hat c(r)^2}{\hat g(r)^2} dt^2 +dx^2+dy^2+ \hat g(r)^2 dr^2\right)\,,
\]
and non-zero components
of the gauge potential,
\[
\label{gaugepotential}
A_t = \frac{e L}{\kappa} \frac{\hat c(r)\hat a(r)}{r \hat g(r)} \;, \qquad
A_y = \frac{e L}{\kappa} \hat B x \;.
\]
This conveniently leads to simple expressions for the local chemical potential and magnetic field,
\[
\mu_{\rm loc}(r)=\frac{e}{\kappa}\hat a(r) \,, \qquad 
H_{\rm loc}(r)=\frac{e}{\kappa L}\hat B r^2 \,.
\]
Inserting this ansatz into the Einstein-Maxwell field equations 
coupled to a charged fluid leads to a system of ODE's for $\left\{\hat a(r), \hat c(r), \hat g(r)\right\}$,
which can be solved by numerical methods along the lines of~\cite{Puletti:2010de}.
The numerical algorithm is implemented on dimensionless field variables, denoted by a hat, 
obtained from their dimensionful counterparts by absorbing appropriate powers of $\kappa$, 
$e$, and $L$. For more detail, we refer to the Appendix. 

The local chemical potential $\mu_{\rm loc}$ vanishes both at the horizon, $r=1$, and at the AdS 
boundary, $r=0$. It follows that $\ell_{\rm filled}$ can only be positive inside a finite range 
$r_i > r > r_e$, defining the radial region where the fluid is supported. The region between the
horizon and the inner edge of the fluid $1>r>r_i$ is described by a vacuum dyonic black brane solution.
We also have a vacuum solution in the region outside the fluid $r_e>r>0$, but with different black brane 
parameters due to the additional mass and charge of the intervening fluid.  
The magnetic field is the same in all three regions as the fluid particles
do not carry magnetic charge, but the bulk magnetization varies due to the presence of the fluid.

\begin{figure}[h]
\begin{center}
\subfigure{\includegraphics[width=0.3\paperwidth]{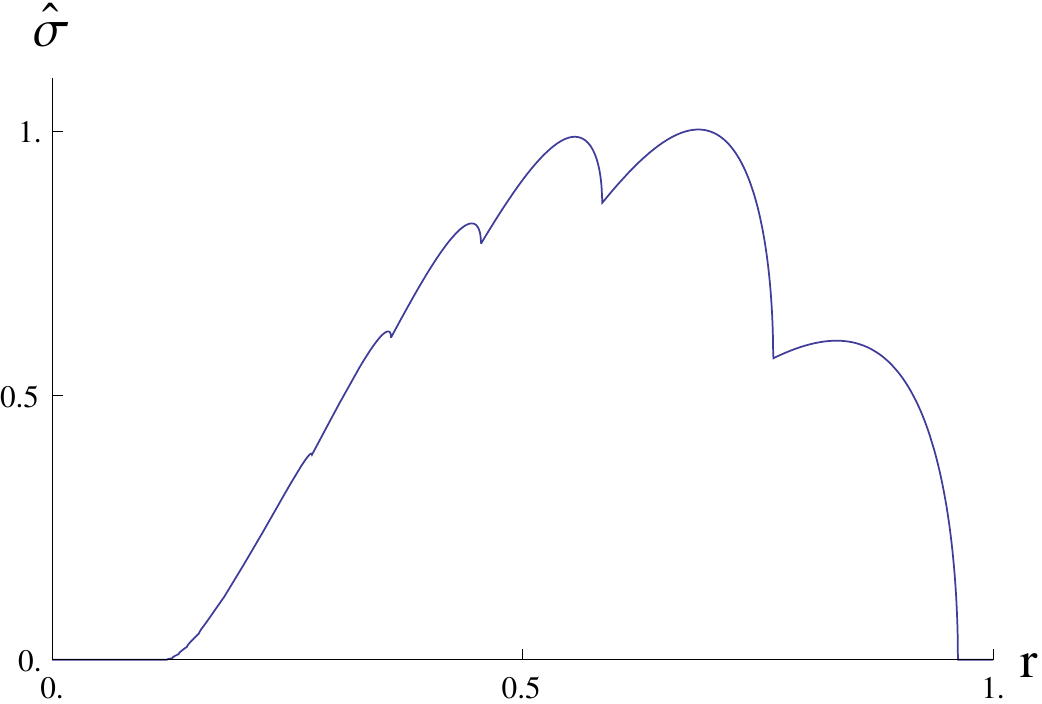}\label{fig:sigmaplot}}\qquad
\caption{
Profile of the fluid charge density $\hat{\sigma}$, with
parameters chosen so that $\max \ell_{\rm filled} = 5$.
}
\label{fig:fluid_profile_LL}
\end{center}
\end{figure}

A typical profile for the fluid charge density $\hat{\sigma}$
is shown in fig.~\ref{fig:fluid_profile_LL}. 
The fluid does not extend all the way to the horizon or the boundary and the bump-like shape of the 
profile reveals the presence of jumps in the local number of filled levels.

The AdS/CFT dictionary relates thermodynamic quantities of the dual field theory to
properties of the bulk metric and gauge field. Temperature $\hat{\mathcal{T}}$ 
and entropy $\hat{\mathcal{S}}$ in the dual field theory are
given by the Hawking temperature and entropy of the black brane, while the 
energy $\hat{\mathcal{E}}$, 
chemical potential $\hat{\mu}$, charge $\hat{\mathcal{Q}}$,
magnetic field strength $\hat{\mathcal{B}}$ and magnetization $\hat{\mathcal{M}}$
of the boundary dual are read off from the asymptotic behavior of the bulk fields. 
The free energy follows from evaluating the on-shell action, and 
is found to satisfy the standard thermodynamic relation, 
\be
\hat{\mathcal{F}} &=& \hat{\mathcal{E}} - \hat{\mathcal{S}} \hat{\mathcal{T}} - \hat{\mu} \hat{\mathcal{Q}} \; .
\ee
The field equations of the bulk theory give rise to the following equation of state for the dual theory
(see Appendix for details),
\be
\frac{3}{2}\hat{\mathcal{E}}
	&=&	\hat{\mathcal{S}} \hat{\mathcal{T}} + \hat{\mu} \hat{\mathcal{Q}} - \hat{\mathcal{M}} \hat{\mathcal{B}}	\; ,
\ee
which agrees with the corresponding relation for dyonic black branes in $AdS_4$. 

\paragraph{Results:}
The electron fluid solution is only found for restricted values of $\hat{\mathcal{B}}$ and $\hat{\mathcal{T}}$.
In the absence of a magnetic field it was already observed in~\cite{Puletti:2010de,Hartnoll:2010ik}, that there 
is a critical temperature $\hat{\mathcal{T}}_c$ above which there is no electron fluid and a phase transition to 
a vacuum black brane configuration occurs.
A non-zero magnetic field brings two new aspects.
First of all, the transition temperature goes down monotonically as $\hat{\mathcal{B}}$ is
increased until it reaches zero at a critical magnetic field $\hat{\mathcal{B}}_c$,
above which no electron fluid is supported at any temperature.
This is evident from our numerical solutions, but can also be inferred from the analytic dyonic 
black brane solution (provided in the Appendix).
Raising either $\hat{\mathcal{B}}$ or $\hat{\mathcal{T}}$ lowers the maximum value reached by the local
chemical potential $\mu_{\rm loc}$ as a function of $r$ in the dyonic black brane background. 
For sufficiently high $\hat{\mathcal{B}}$ and/or $\hat{\mathcal{T}}$ the number of occupied levels 
$\ell_{\rm filled}$ in \eqref{levelcondition} will be nowhere positive so that no fluid can be supported 
and the vacuum dyonic black brane is the only available solution. 

Second, the order of the phase transition changes.
Whereas in~\cite{Puletti:2010de,Hartnoll:2010ik} it was found to be of third order,
it becomes second order in the presence of a magnetic field. This can be shown
by a similar analytic argument as was used in~\cite{Hartnoll:2010ik} for the $\hat{\mathcal{B}}=0$ case.
Consider a temperature just below the transition temperature at $\hat{\mathcal{B}}\neq 0$, keeping the 
magnetic field fixed. The condition $\hat\mu_{\rm loc} > \hat m$ is then satisfied in a narrow band 
in the radial direction in the dyonic black brane solution and inside this band the 
fermions can occupy the lowest Landau level only. Furthermore, the back-reaction on the 
geometry due to the fermion fluid can be neglected at temperatures very close to the 
transition. In the presence of the fermion fluid the free energy of the system is lowered 
by an amount given by the on-shell action of the fluid, {\it i.e.} the integral of the fluid 
pressure. Near the phase transition the pressure, given by equation \eqref{pdef} in the Appendix,
scales as $\hat p \propto (\hat{\mu}-\hat{m})^{3/2}$. A short calculation along the lines
of~\cite{Hartnoll:2010ik} then results in a free energy difference, 
$\Delta \hat{\mathcal F}\propto (\hat{\mathcal{T}}_c-\hat{\mathcal{T}})^2$, 
between the solution with a fluid and a vacuum 
dyonic black brane, indicating a second order phase transition. The same behavior is also seen 
in our numerical solutions of the field equations with the full back-reaction included. In the 
limit of vanishing magnetic field, the different Landau levels collapse to a continuum and 
one obtains a softer dependence, $\hat p \propto (\hat{\mu}-\hat{m})^{5/2}$, that leads to a
third order phase transition as was found in~\cite{Hartnoll:2010ik}.

At non-vanishing magnetic field one can also approach the phase transition by varying
$\hat B$ at fixed temperature. In this case we find 
$\Delta \hat{\mathcal F}\propto (\hat{\mathcal B}_c-\hat{\mathcal{B}})^2$
and the phase transition is again of second order. We anticipate that going beyond the 
Thomas -- Fermi approximation will change the nature of the phase transition. Indeed, a first 
order phase transition was found at $\hat{\mathcal{B}}=0$ using WBK wave functions for the
fermions in~\cite{Medvedyeva:2013rpa} and we expect this would be the case at finite 
$\hat{\mathcal{B}}$ as well.

\begin{figure}[h]
\begin{center}
\includegraphics[width=0.3\paperwidth]{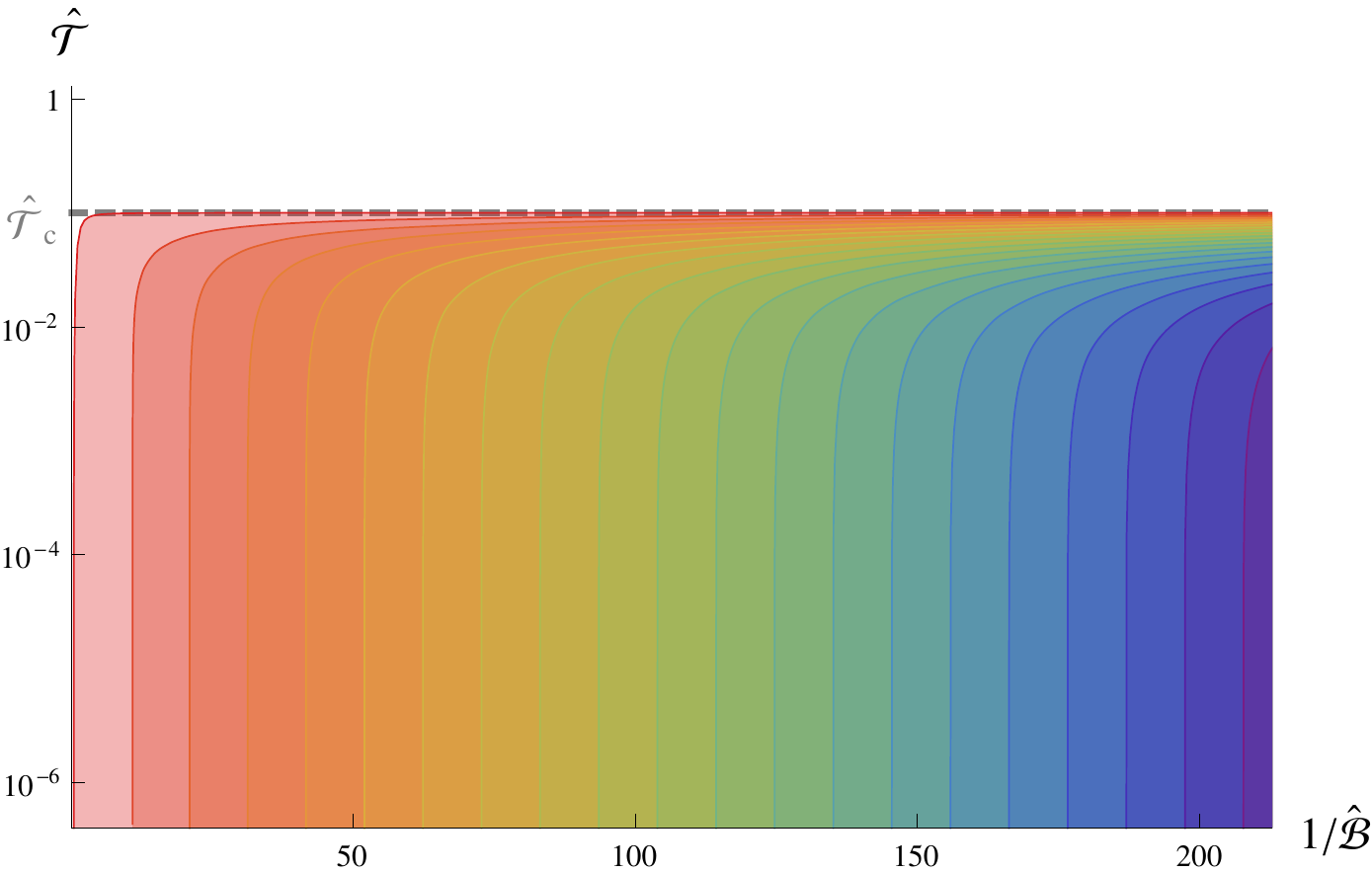}
\end{center}
\caption{Phase diagram of dyonic electron fluid solutions for $\hat m=0.5$, with axes normalized 
such that the boundary chemical potential is $\hat \mu = 1$. The different color shadings mark parameter
regions with different maximum numbers of occupied Landau levels, increasing from left to right.
In the limit $\hat{\mathcal{B}} \to 0$ the edges between the regions all asymptote to 
$\hat{\mathcal{T}}_c$, the maximal temperature at which the electron fluid is supported at 
$\hat{\mathcal{B}}=0$.
}
\label{fig:phase_diagram}
\end{figure}
The $\hat{\mathcal{B}} - \hat{\mathcal{T}}$ phase diagram reveals a periodic feature in $1/\hat{\mathcal{B}}$.
A representative plot for $\hat m=0.5$ is displayed in fig.~\ref{fig:phase_diagram}.
Changing the value of $\hat m$ in the numerical calculations does not significantly affect the phase 
diagram, apart from changing the critical values on the axes.
Different colors mark different values of the maximum value of filled Landau levels $\ell_{\rm filled}$,
which increases from left to right.
At low temperatures the edges between regions with a different number of occupied levels occur at 
equal intervals in $1 / \hat{\mathcal{B}}$.
This periodic feature is even more apparent in the plots showing the magnetization $\hat{\mathcal{M}}$ 
as a function of $\hat{\mathcal{B}}$ in fig.~\ref{fig:MvsBplot}.
For temperatures close to the critical transition temperature $\hat{\mathcal{T}}_c$,
the magnetization differs only slightly from that of a dyonic black brane at the same temperature and magnetic field strength but when the temperature is lowered the magnetization oscillates.
The oscillations are clearly visible when $\hat{\mathcal{B}} \gg \hat{\mathcal{T}}$, which is 
the regime where the de Haas -- van Alphen effect is observed experimentally.

\begin{widetext}
{\onecolumngrid
\begin{figure}[h!]
\begin{center}
\hspace{-0.01\paperwidth}
\includegraphics[width=0.265\paperwidth]{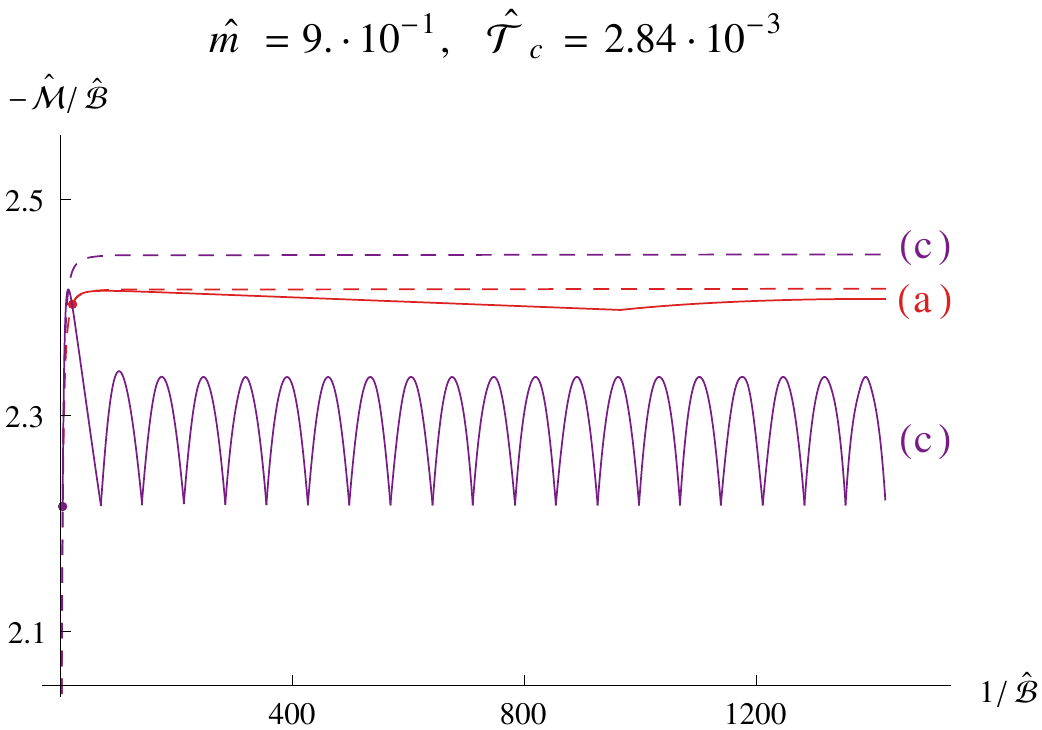}\quad
\includegraphics[width=0.265\paperwidth]{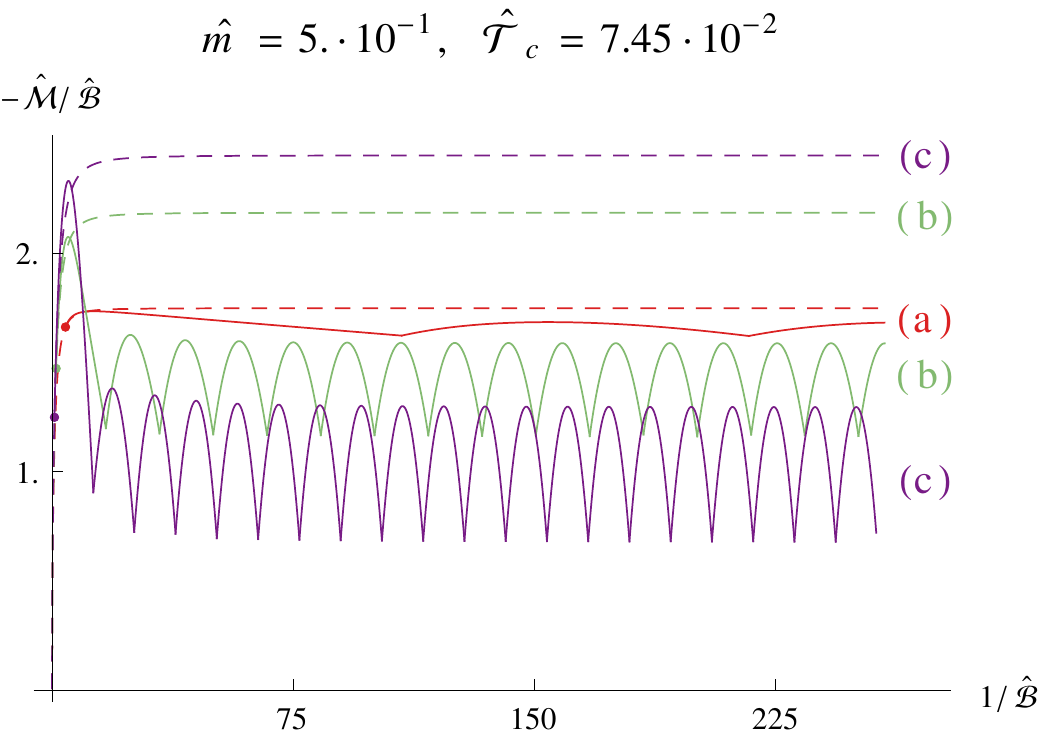}\quad
\includegraphics[width=0.265\paperwidth]{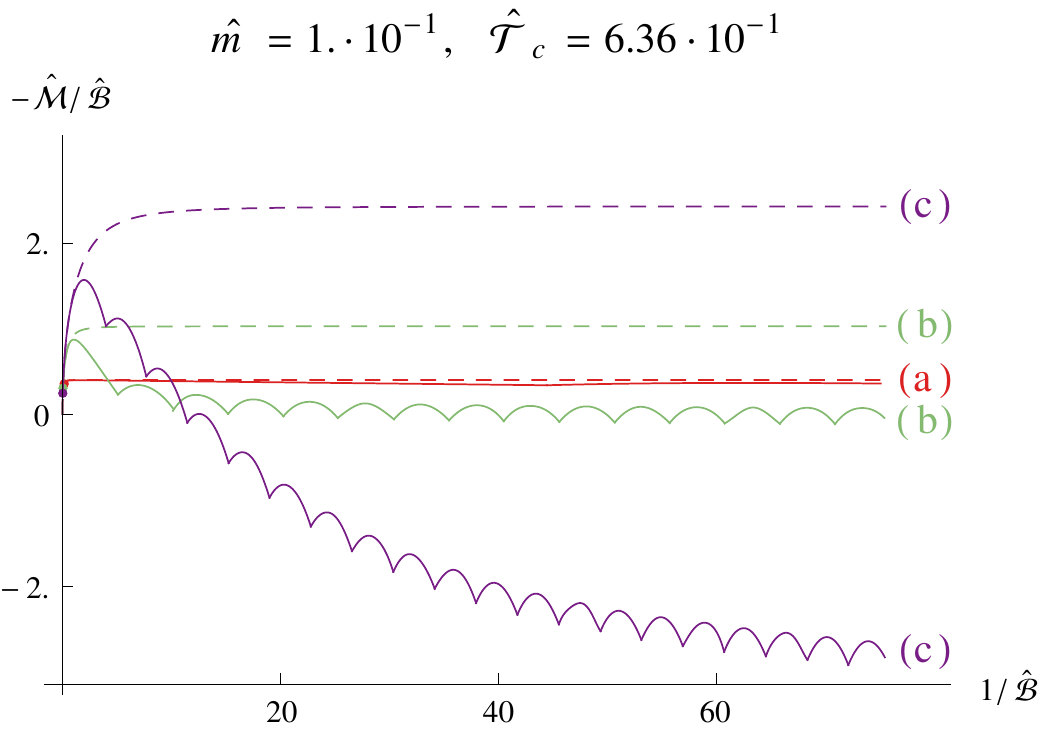}
\end{center}
\caption{
Magnetization vs. magnetic field strength for various values of $\hat m$.
Solid lines correspond to the electron fluid geometry,
dashed ones to a dyonic black brane at same temperature and chemical potential.
The labels denote temperatures $\hat{\mathcal T}/\hat{\mathcal T}_c = 0.9\;(a), \,0.3\;(b), \,3 \cdot 10^{-3}\;(c)$.
In the leftmost plot $(b)$ was omitted due to too much visual
overlap with the other curves. In the rightmost plot the relative sign of $\hat{\cal{M}}$ and $\hat B$
changes as a function of magnetic field, indicating a crossover from a diamagnetic to a paramagnetic state.}
\label{fig:MvsBplot}
\end{figure}
}
\end{widetext}

Another feature is that the magnetization of the electron fluid configuration is 
lower than that of a dyonic black brane with the same parameters.
This is more pronounced as the value of $\hat m$ is lowered and when $\hat m$ becomes small enough, the state 
crosses over from diamagnetic to paramagnetic.
The local magnetization is the sum of two contributions,
a diamagnetic one which originates from the black brane 
and a paramagnetic one due to the fluid which is a gas of free electrons 
at zero temperature. Varying the parameter $\hat m$
tunes the gravitational attraction between the black brane in the 
center and the electron fluid surrounding it. The weaker the interaction, the larger the 
fluid region can grow and the more dominant the paramagnetism becomes.

For small values of the magnetic field, the overall amplitude of the magnetization is
linear in $\hat{\mathcal{B}}$, as can be seen in fig.~\ref{fig:MvsBplot}.
This differs from the behavior predicted by the Landau theory of Fermi liquids via
the Kosevich~--~Lifshitz formula~\cite{Lifshitz-Kosevich,Onsager}.
It also differs from earlier holographic results obtained in a probe limit in \cite{Hartnoll:2010xj, Blake:2012tp} 
and appears to be due to the gravitational back-reaction which is included in our model. 
\begin{figure}[h]
\begin{center}
\includegraphics[width=0.3\paperwidth]{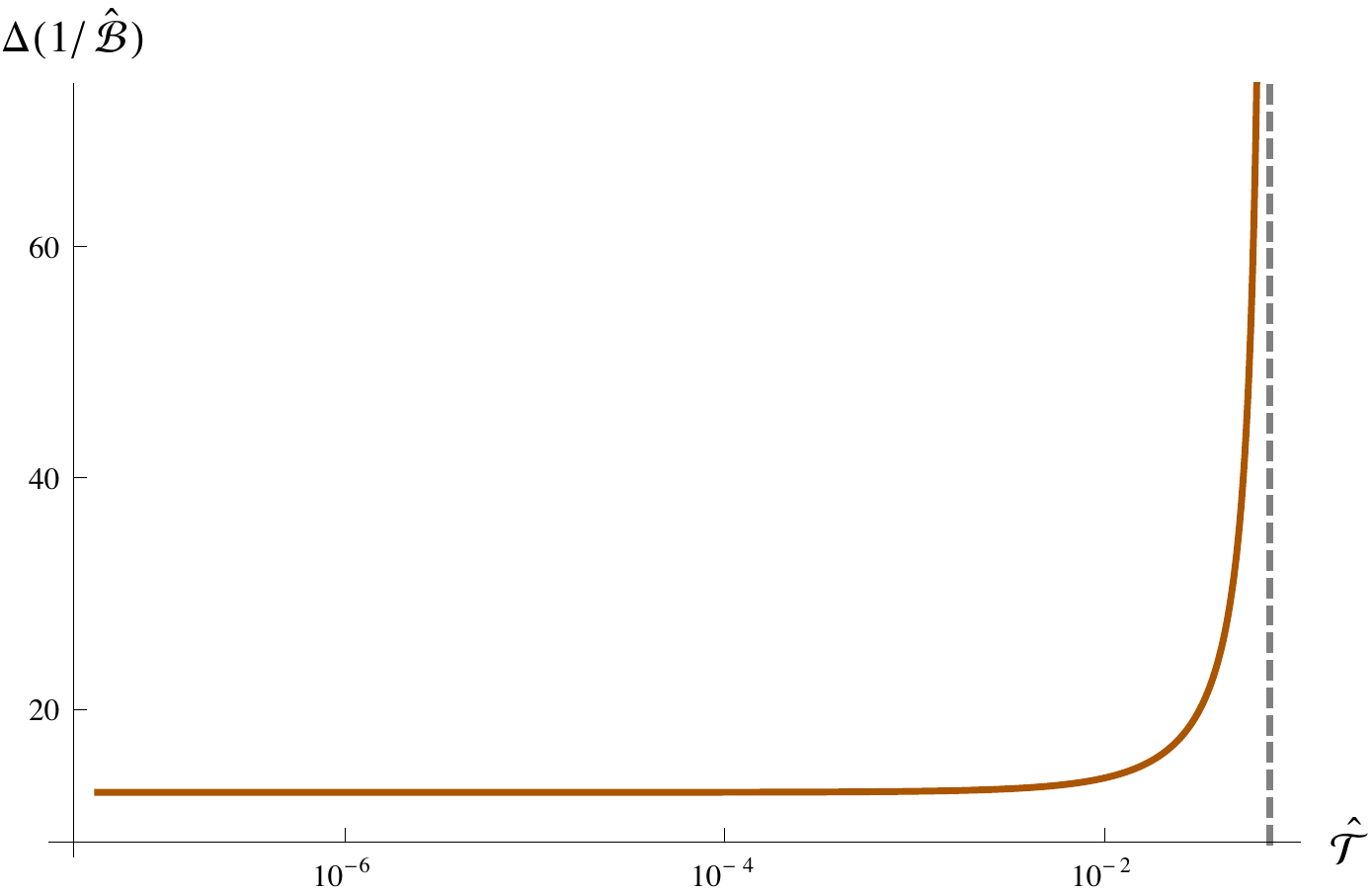}
\caption{
Period of the de Haas -- van Alphen oscillations as a function of temperature for $\hat{m} = 0.5$, 
the dotted vertical line marks the critical temperature where the solution makes the transition to a 
dyonic black brane. 
}
\label{fig:deltaB}
\end{center}
\end{figure}

The plots in figure \ref{fig:MvsBplot} do not show any overlap of oscillations with
different periods, suggesting that a single Fermi surface is responsible for the phenomenon. 
This is in agreement with \cite{Hartnoll:2010xj, Hartnoll:2011dm}, where it was argued that 
magnetic oscillations are dominated by a single (extremal) Fermi momentum, despite the 
large number of holographically smeared Fermi surfaces in this system,
which turn into a continuum in the Thomas -- Fermi limit.

Figure~\ref{fig:deltaB} shows the period in $1/{\hat{\mathcal B}}$ of the quantum oscillations 
{\it vs.}\ temperature. It reveals a similar trend as is observed in experiments,
where the period is constant at low temperature but increases with rising $T$
until the oscillations get washed out at higher temperatures. Our numerical 
results suggest that the oscillation period diverges in the holographic model
at the critical temperature for the transition to the dyonic black brane, above 
which the electron fluid is no longer supported.

\paragraph{Discussion:}
We have presented a holographic model for a 2+1 dimensional system of strongly correlated electrons 
in a magnetic field, involving 3+1 dimensional fermions treated in a Thomas -- Fermi approximation in an
asymptotically AdS dyonic black brane background, taking into account both the gravitational and 
electromagnetic back-reaction due to the charged matter. The system exhibits de Haas -- van Alphen 
oscillations that appear to be dominated by a single sharp Fermi surface, 
while the oscillation amplitude has a non-Fermi liquid character that departs from earlier probe 
fermion computations. 

While the semi-classical model studied here provides a relatively simple framework for numerical 
computations, it is rather crude. A Thomas -- Fermi treatment of bulk fermions is known to wash out 
some quantum features that are present in more realistic models~\cite{Puletti:2011pr, Hartnoll:2011dm}
and the sharp edges of the fluid profile for individual Landau levels can introduce fictitious non-analyticities
into observables that involve derivatives acting on the bulk fields~\cite{Hartnoll:2010ik}. The latter 
problem can presumably be remedied by introducing thermal effects in the bulk fluid or by replacing the 
anisotropic electron star by a quantum many-body 
model based on WKB wave functions for bulk Landau levels, along the lines of~\cite{Medvedyeva:2013rpa}, 
where the tails of the fermion wave functions naturally smooth out the edges found in the fluid description,
but we leave this for future work. 


\begin{acknowledgments}
\paragraph{Acknowledgements:}

We would like to thank N. Bucciantini, E. Kiritsis, V. Jacobs, K. Schalm, and J. Zaanen
for helpful discussions. 
This work was supported in part by the Icelandic Research Fund and by the University of 
Iceland Research Fund. V.G.M.P. and L.T. acknowledge the Swedish Research Council for 
funding under contracts 623-2011-1186 and 621-2014-5838,
respectively.
T.Z. was partially supported by the Nederlandse Organisatie voor Wetenschappelijk
Onderzoek (NWO) under the research program of the Stichting voor Fundamenteel Onderzoek
der Materie (FOM).
\end{acknowledgments}

\begin{appendix}

\subsection{Appendix}
\label{app:appendix}


\paragraph{Field equations:}
We adapt the electron star construction 
developed in~\cite{Hartnoll:2009ns,Hartnoll:2010gu,Puletti:2010de,Hartnoll:2010ik} to include
the effects of a background magnetic field.
Our starting point is the action (1) in the main text, which describes a charged fluid coupled to
Einstein -- Maxwell theory. 
We will be considering static solutions describing
a charged fluid suspended above the horizon of a planar dyonic black brane in 3+1 dimensional
asymptotically AdS spacetime with 
radial electric and magnetic fields and translation symmetry in the two transverse directions. 
The field equations are given by
\be\nn
R_{\mu \nu} -\half R\, g_{\mu\nu}+ \Lambda \, g_{\mu \nu}
	&=&	\kappa^2\left(T_{\mu \nu}^{\rm em}+ T_{\mu \nu}^{\rm fl}+T_{\mu \nu}^{J}+ T_{\mu \nu}^{M}\right)	\, ,	\\
 \nabla_\nu F^{ \mu \nu} &=&  e^2 \left(J^\mu+\nabla_\nu  M^{\mu\nu}\right)	\, ,	
\label{eom_general} 
\ee
where $J^\mu$ and $M^{\mu\nu}$ are the fluid current and magnetization tensor,
\be
J^\mu = - \frac{\delta \CL_{\rm fl}}{\delta A_\mu}\,, \qquad
M^{\mu\nu}=-2\, \frac{\delta \CL_{\rm fl}}{\delta F_{\mu\nu}}\,,
\ee
and the stress energy tensors are given by
\be
T_{\mu \nu}^{\rm em}  &=&\frac{1}{e^2}\Big( F_{\mu\l} F_{\nu}^{~~\l} - \frac{1}{4} g_{\mu\nu} F_{\l\s} F^{\l\s} \Big)\,, \nn 
\\ \nn
  T_{\mu \nu}^{\rm fl} &=& - g_{\mu\nu} \CL_{\rm fl}\,,\\ \nn
 T_{\mu \nu}^{J} &=&
 -J_{(\mu}  A_{\nu)} +u^\l A_{\l}u_{(\mu} J_{\nu)} - u^\l J_\l u_{(\mu}  A_{\nu)}  
 \,,\\ 
 T_{\mu \nu}^{M} &=& M_{\l (\mu}F_{\nu)}^{~\l} +u^\l F_{\l}^{~\rho} u_{(\mu} M_{\nu)\rho} -u^\l M_{\l \rho} u_{(\mu} F_{\nu)}^{~~\rho}    
 \, .~~~~~
 \label{def_tensors}
\ee
Let $e_\mu^A$ denote a local Lorentz frame where the fluid is at rest. The fluid four velocity is 
then given by $u^\mu=e^\mu_0$ and 
the fluid components experience a local chemical potential and a local magnetic field, 
\be
\label{localvariables}
\mu_{\rm loc}=A_\mu u^\mu\,, \qquad 
H_{\rm loc}= \, e_{\>1}^{[\mu} e_{\,2}^{\nu]} F_{\mu \nu} \, ,
\ee
which completely determine the state of the charged fluid at a given point in the bulk spacetime. 
In particular, the electric current and magnetic polarization, which can be expressed in terms 
of local charge and magnetization densities, 
$J^\mu=\sigma u^\mu$, $M^{\mu \nu}=2\, \eta\, e_{\>1}^{[\mu} e_{\,2}^{\nu]}$,
and the on-shell fluid Lagrangian density, given by the pressure $\CL_{\rm fl}= - p$ for the static solutions we 
are considering, are all functions of $\mu_{\rm loc}$ and $H_{\rm loc}$. 
 
The formalism we are using derives from so called spin fluid models, which have been studied in general 
relativity since the 1970's~\cite{1970PhRvD...2.2762S,Ray:1972:LDP,BaileyIsrael,deOliveira:1991ja,Brown:1992kc,1995PhLA..201..381D}. We have only presented the minimal ingredients needed to describe the static
geometries that are of interest here, but the full formalism can also handle more general dynamical backgrounds. 

\paragraph{Fluid variables:}
The bulk variables that describe the fluid are the charge density $\sigma$, 
the magnetization density $\eta$ and the pressure $p$.
The fluid components are locally free fermions in an external magnetic field along the radial direction, 
with dispersion relation
\be
\label{disp_rel}
E_\ell^2=	m^2 + k^2 + \Big(2\ell +1\Big) \gamma H_{\rm loc} \pm \gamma  H_{\rm loc} \,,	
\ee
where the index $ \ell \geq 0$ labels the Landau levels, $\gamma$ is a constant proportional to the
gyromagnetic ratio of the constituent fermions, and the $\pm$ in the rightmost term is due to Zeeman splitting.
There is a degeneracy between different sign Zeeman states in adjacent Landau levels. The sum
over levels can therefore be rearranged into a sum $\sum'_{\ell \geq 0}$, where the prime indicates
inserting a relative factor of $1/2$ in the $\ell=0$ term.
The density of states is
\be
\label{density_states}
n(E) & =&
{{\beta}\gamma}{H}_{\rm loc} 
         {\sum_{\ell \geq 0}}' \theta(E^2 - \epsilon_\ell^2) \frac{E}{\sqrt{E^2 - \epsilon_\ell^2}}\,,
\ee
where $\beta$ is a constant and $\epsilon_\ell = \sqrt{ m^2 + 2 \, \ell \, \gamma\, H_{\mathrm{loc}} }$ 
is the energy in the Landau level labelled by $\ell$. In the limit of weak magnetic field the sum 
over Landau levels can be replaced by an integral, which is easily performed to reproduce
the density of states used to construct an electron star in zero magnetic field in \cite{Hartnoll:2010gu}.

The local charge density $\s$ is obtained from the density of states via
\be
\s = \int_{0}^{\mu_{\rm loc}} n(E) dE \,,
\ee
and the pressure and magnetization density are obtained from the charge density by the
thermodynamic relations,   
\be
\frac{\partial p}{\partial \mu_{\rm loc}} = \sigma \,, \quad
\frac{\partial p}{\partial H_{\rm loc}} = \eta \,, \qquad \label{thermorelations}
\ee
analogous to the electron star~\cite{Hartnoll:2010gu}.
The constant of integration in $p$ is fixed such that $p$ vanishes for $\sigma=0$. 
Using the density of states in \eqref{density_states} leads to the following explicit expressions for the 
fluid variables in terms of the local chemical potential and local magnetic field,
\begin{widetext}
{\onecolumngrid
\be
\label{pressure_scaling}
\s &=&
{{\beta}\gamma}{H}_{\rm loc} 
         {\sum_{\ell \geq 0}}' {\theta\left(\mu^2_{\rm loc} - \epsilon_\ell^2\right) \sqrt{\mu^2_{\rm loc} - \epsilon_\ell^2}}\,, \\ \label{pdef}
 p&=&
 \frac{\gamma \beta}{2}  H_{\rm loc} {\sum_{\ell \geq 0}}'  \theta\left(\mu^2_{\rm loc} - \epsilon_\ell^2  \right) 
	\left[
 \mu_{\rm loc}  \sqrt{\mu_{\rm loc}^2 - \epsilon_\ell^2 }
 - \epsilon_\ell^2 \log \left(\frac{\sqrt{\mu_{\rm loc}^2 -\epsilon_\ell^2 }+\mu_{\rm loc} }{\epsilon_\ell}\right)
   \right]\,,
   \\
 \eta &=& 
  \frac{\gamma \beta }{2} {\sum_{\ell \geq 0}}' \theta\left(\mu^2_{\rm loc} - \epsilon_\ell^2  \right)
 \left[\mu_{\rm loc} \sqrt{\mu_{\rm loc}^2 - \epsilon_\ell^2 }-\left( 2\epsilon_\ell^2 - m^2 \right) \log \left(\frac{\sqrt{\mu_{\rm loc}^2 - \epsilon_\ell^2}+\mu_{\rm loc}}{\epsilon_\ell}\right)\right]\,. 
\ee
}\end{widetext}

The matter stress-energy tensors in \eqref{def_tensors} reduce to
\be\nn
\label{def_tensors_exp}
&& T^{\rm fl}_{\mu\nu}= p \, g_{\mu\nu}\,, \qquad
T_{\mu\nu}^J = \mu_{\rm loc} \,\sigma \,u_\mu u_\nu \,, 
  \\
&& T^M_{\mu\nu}=-\frac{1}{2}H_{\rm loc}\, \eta \, (e_\mu^1 e_\nu^1+e_\mu^2 e_\nu^2) \,.
\ee
We proceed to solve the 
combined Einstein and Maxwell equations for this system
with the above expressions for the fluid variables.

\paragraph{Metric and gauge field ansatz:}
Parameterizing non-vanishing components of the tetrad as
\be
\label{tetrad} 
e_t^0 = \frac{L}{r} \frac{\hat c(r)}{\hat g(r)} \;, \quad
e_x^1 = e_y^2	= \frac{L}{r} \;, \quad
e_r^3 = \frac{L}{r} \hat g(r) \;,
\ee
and the gauge potential as
\be
\label{gaugepotential}
A_t = \frac{e L}{\kappa} \frac{\hat c(r)\hat a(r)}{r \hat g(r)} \;, \qquad
A_y = \frac{e L}{\kappa}\hat B x \;,
\ee
yields the following local chemical potential and magnetic field,
\be
\mu_{\rm loc}(r)=\frac{e}{\kappa}\hat a(r) \,, \qquad 
H_{\rm loc}(r)=\frac{e}{\kappa L}\hat B r^2 \,.
\ee
Hats denote dimensionless quantities and we find it useful to convert all parameters 
and field variables into dimensionless form~\cite{Hartnoll:2010gu},
\be\nn
\label{rescale_parameters}
&& \hat m =\frac{\kappa}{e}m \,,\quad
\hat\beta =\frac{e^4L^2}{\kappa^2}\beta \,, \quad
\hat\gamma=\frac{\kappa}{eL} \gamma\,,
\\ \nn
&& \hat\mu_{\rm loc}=\frac{\kappa}{e} \mu_{\rm loc}\,, \quad
\hat H_{\rm loc}=\frac{\kappa L}{e} H_{\rm loc}\,, \quad
\\ \label{rescale_variables}
&& \hat \sigma=e\kappa L^2\, \sigma \,, \quad
\hat p =\kappa^2L^2\, p\,, \quad
\hat\eta = e\kappa L\, \eta \,.
\ee

\paragraph{Final form of the field equations:}
The Einstein and Maxwell equations~\eqref{eom_general}, with stress-energy tensors given by 
\eqref{def_tensors_exp} and the ansatz \eqref{tetrad}-\eqref{gaugepotential} for the metric and the 
Maxwell gauge field, reduce to a system of first order ordinary differential equations, 
\be
\label{eom_tobias}
\hspace{-3mm} r \frac{d\hat c}{dr}&=&-\frac{1}{2} \hat c \, \hat g^2 \hat a \, \hat \sigma \, ,\label{eq:eom_1} \\ 
\hspace{-3mm} r \frac{d\hat g}{dr}&=&- \frac{3}{2}\hat g - \frac{1}{4}\hat g^3 (\hat{B}^2 r^4{+}\hat q^2 r^4 {-}6{-}2 \hat p
+2 \hat a\, \hat \sigma ) ,\\ 
\hspace{-3mm} r \frac{d\hat a}{dr}&=&- \frac{\hat a}{2} - r^2 \hat g\,\hat q - \frac{\hat a\, \hat g^2}{4} 
(\hat B^2 r^4 {+}\hat q^2 r^4 {-}6 {-}2 \hat p) ,\\ 
\hspace{-3mm} r \frac{d\hat q}{dr}&=&-\frac{1}{r^2}\,\hat g\, \hat \sigma \,, \\ 
\hspace{-3mm} r\, \frac{d\hat{\mathfrak{M}}}{dr}
&=&\hat B \, r -\frac{\hat\eta}{r} + \half\hat a \, \hat g^2 \, \hat \sigma\, \hat{\mathfrak{M}}  \,.
\label{eq:eom_5}
\ee
We have introduced two auxiliary functions. One is $\hat q(r)$, which is related to the value of the local electric field 
by $e_0^te_3^rF_{tr}=\frac{e}{\kappa L}r^2\hat q$. The other is $\hat{\mathfrak{M}}$, which originates from the functional derivative  $\frac{\delta S}{\delta F_{xy}}$  of the on-shell action, and whose value at $r=0$ is the magnetization in the boundary theory, $\hat{\mathcal{M}}= \lim_{r\to 0}\hat{\mathfrak{M}}(r)$.
Finally, energy-momentum conservation can be expressed as
\be
\frac{d\hat p}{dr}=\hat\sigma \frac{d\hat a}{dr}+2 \, r\hat B\, \hat\eta \,. \label{eq:eom_6}
\ee
The following quantity 
%
\be 
\mathcal{\hat Y}
&=&	\hat c \left[\frac{3+\hat p}{r^3}-\frac{3}{r^3 \hat g^2}-\frac{2\hat a \hat  q}{r \hat g}
+2\hat B \, \hat{\mathfrak{M}}-\frac{r(\hat B^2+\hat q^2)}{2} \right] \; \label{eq:constant_Y}~~~~
\ee
is constant along the radial direction $r$ when the field equations are satisfied, and later on we use this to 
determine \eqref{thermo2}, the equation of state in the dual boundary field theory. 

\paragraph{Solutions:}
In the presence of a charged fluid we have to solve the field equations \eqref{eom_tobias}-\eqref{eq:eom_5}
numerically. As discussed in the main text, the fluid is only supported where the local chemical potential 
is larger than the minimum energy state in the lowest Landau level. In solutions with a non-vanishing fluid
profile this condition is met inside a radial range $1>r_i>r> r_e>0$, where $r=1$ is the radial location of the brane horizon, and $r=0$ marks spatial infinity of the spacetime.

In the region $1> r>r_i$, there is no fluid and the solution of the field equations is a dyonic black brane,
\be\nn
&& \hat c(r)=1\,,\quad 
\frac{\hat a(r)}{r\hat g(r)}=\hat Q (1-r) \,, \quad
\\ 
&& \frac{1}{\hat g(r)^2}	=1- \frac{2+\hat Q^2+ \hat B^2}{2}r^3 
+ \frac{\hat Q^2+\hat B^2}{2} r^4\,. 
\label{interiorbrane}
\ee

The local chemical potential grows as we move outwards from the horizon and by equation (5) in the 
main text the lowest Landau level can be occupied if $\hat\mu_\textrm{loc}^2\geq\hat m^2$. The radial 
position $r=r_i$ outside the black brane horizon where this condition is first satisfied defines the 
inner edge of the fluid. The dyonic black brane solution then provides initial data at $r=r_i$ for the 
subsequent numerical evaluation of the system of equations \eqref{eq:eom_1}-\eqref{eq:eom_5} in 
the fluid region. At high temperature, the condition is not satisfied anywhere outside the brane horizon.
In this case there will be no fluid and the only solution is the dyonic black brane itself. Whenever a 
solution with a fluid present exists, however, it has a lower free energy density than a dyonic black 
brane at the same temperature. 

Returning to the description of the fluid solution, the local chemical potential reaches a maximum and 
then decreases towards the exterior edge $r=r_e$ where the fluid is no longer supported. Outside the 
fluid, we have another dyonic black brane solution
with parameters determined by the output of the numerical integration at $r=r_e$, 
\be
\label{exteriorbrane} \nn
\hat c(r)&=&\mathfrak{c} \,,\quad 
\frac{\hat a(r)}{r\hat g(r)}=\hat{\mu} - \hat{\mathcal{Q}} \, r ,\quad
\\ 
 \frac{1}{\hat g(r)^2}	&=&1- \hat{\mathcal{E}}\, r^3
 + \frac{\hat{\mathcal{Q}}^2+\hat{\mathcal{B}}^2}{2}  r^4\,, 
\ee
with
\be
\hat{\mathcal{B}}&=&\hat B\,,~~ \mathfrak{c}=\hat c(r_e) \,,~~
\hat{\mathcal{Q}}=\hat q(r_e)  \,, ~~  \hat\mu = \frac{\hat a(r_e)}{r_e \hat g(r_e)}+r_e \hat q(r_e) \nonumber \\
\hat{\mathcal{E}}&=& \frac{1} {r_e^3}\left(1-\frac{1}{\hat g(r_e)^2} \right)
+\frac{r_e}{2}\left(\hat B^2+\hat q(r_e)^2\right) \,. 
\ee
Calligraphic letters denote boundary quantities and hatted calligraphic letters denote 
dimensionless boundary quantities.
The boundary magnetization is given by the value of the auxiliary function $\hat{\mathfrak{M}}(r)$ at $r=0$, 
which can obtained by integrating \eqref{eq:eom_5} from the outer edge of the fluid to the AdS boundary, 
\be
\hat{\mathcal{M}}= \lim_{r\to 0}\hat{\mathfrak{M}}(r)=-\hat B \, r_e+ \hat{\mathfrak{M}}(r_e)\,.
\ee
\paragraph{Thermodynamics:}
The dual field theory temperature and entropy are the
Hawking temperature and Bekenstein-Hawking entropy of the black brane.
Restoring dimensions to our quantities, we obtain
\be
{\mathcal{T}} = \frac{1}{8\pi \mathfrak{c} L}
\left( 6 - \hat{Q}^2 - \hat{B}^2 \right) \, ,\qquad 
{\mathcal{S}} =\frac{2\pi}{\kappa^2}V_2\,,~~~~~~
\ee
where the $\mathfrak{c}$ keeps track of the different time normalization at the inner and outer edges 
of the fluid
and $V_2$ is the volume of the two-dimensional boundary. 
The free energy is computed from the on-shell regularized Euclidean action, 
\be
\label{thermo1}
\mathcal{F}=\mathcal{E} - \mathcal S \mathcal{T} - \mu \mathcal{Q}\,.
\ee
Evaluating the conserved charge $\mathcal Y$ \eqref{eq:constant_Y} at the horizon of \eqref{eq:constant_Y}, and at the 
boundary \eqref{exteriorbrane} gives the thermodynamic relation
\be
\label{thermo2}
\frac{3}{2}\mathcal{E}
	=	\mathcal{S} \mathcal{T} + \mu \mathcal{Q}-\mathcal{M} \mathcal{B}\,, ~~~
\ee
once we restore dimensions according to
\be\nn
\label{boundary_fe}
&& \mathcal{E}	=	\frac{V_2}{\kappa^2 \, L}\hat{\mathcal{E}}	\,, ~~~
\mu			= \frac{eL}{\kappa} \,  \hat \mu\,, ~~~
\mathcal{Q}=\frac{V_2}{\kappa e L^2} \hat{\mathcal{Q}}\,,
\\ 
&&   \mathcal{B}=\frac{eL}{\kappa}\, \hat{ \mathcal{B}}\,, ~~~
\mathcal{M}= \frac{V_2}{\kappa e L^2} \hat{\mathcal{M}}\,.
\ee
\end{appendix}

\bibliography{AdSCMT.bib}

\end{document}